\documentclass
[12pt]{article}
\usepackage[dvips]{color}
\usepackage{epsfig}
\usepackage{amsmath}
\usepackage{graphicx}
\begin{document}
\title {Baryon Binding Energy in Sakai-Sugimoto Model }
\author{J. Sadeghi $^{a,}$\thanks{Email:
pouriya@ipm.ir}\hspace{1mm}, M. R. Pahlavani$^{a}$\thanks{Email:
m.pahlavani@umz.ac.ir.ir}\hspace{1mm}, S. Heshmatian
$^{a}$\thanks{Email: s.heshmatian@umz.ac.ir}\hspace{1mm} and
and R. Morad$^{a}$\thanks{Email: r.morad@umz.ac.ir}\\
$^a$ {\small {\em  Sciences Faculty, Department of Physics, Mazandaran University,}}\\
{\small {\em P .O .Box 47415-416, Babolsar, Iran}}}
 \maketitle
\begin{abstract}
The binding energy of baryon has been studied in the dual
$AdS_5\times S^5$ string theory with a black hole interior. Here, we
calculate the baryon binding energy in Sakai-Sugimoto model. Also we
check the $T$ dependence of the baryon binding energy. We believe
that this model represents an accurate description of baryons due to
the existence of Chern-Simones coupling with the gauge field on the
brane. We obtain an analytical expression for the baryon binding
energy. Next we plot the baryon binding energy in terms of radial
coordinate. Then by using the binding energy diagram, we determine
the stability range for baryon configuration. And also the position
and energy of the stable equilibrium point is obtained by the
corresponding diagram. Also we plot the baryon binding energy in
terms of temperature and estimate a critical temperature in which the baryon would be dissociated.
\end{abstract}
\newpage
\section{Introduction}
The AdS/CFT correspondence demonstrates a relation between a
conformal field theory in $d$ dimension and a gravitational theory
in $d+1$ dimensional anti de-sitter space [1-7]. An example for this
correspondence is the relation between type IIB string theory in
$AdS_5 \times S^5$ space and $\cal{N}$=4 supersymmetric Yang-Mills
theory on four dimensional boundary of $AdS_5$. Using this
correspondence to calculate the complicated problems of QCD is one
of the more interesting issues nowadays. For example the dynamics of
moving quark in a hot, strongly coupled plasma have been
investigated in [8-17]. Also the jet-quenching parameter of quarks
is one of the interesting properties of the strongly coupled plasma
that there are many calculations to obtain this parameter [17-25].
In Additionally the motion of a quark-antiquark pair have been
studied in [26-31].\\
In the other hand baryons in gauge theory have been studied in
$AdS_5 \times S^5$ dual string theory by introducing the baryon
vertex [2]. In this picture baryons are corresponded to the
configurations which consist of a $D_5 $-brane wrapped on a $S_5 $
and all external quarks are connected to it due to fundamental strings.\\
Baryons may also be studied in Sakai-Sugimoto ($SS$) model [32-35]
with $D_4/D_8/\overline{D_8}$ configuration which presents a
holographic dual of four dimensional QCD with large $N_c$ and
massless flavors. $D_4$-brane is placed on a $S^1$ susy-breaking and
the $D_8/\overline{D_8}$ pairs are transverse to $S^1$. The lower
bound for the radial coordinate $U$ which is transverse to the
D4-branes is $U=U_{kk}$ and the radius of $S^{1}$ diminishes to zero
in this point. The spontaneously symmetry breaking in QCD is
indicated as a smooth interpolation of the $D_8/\overline{D_8}$
pairs in super gravitational background. The $SS$ model suggests
that the solution called skyrmion demonstrations a baryon which is
considered as a $D_4$-brane wrapped on $S^4$. This $D_4$-brane is
the baryon vertex with $N_c$ connected strings. The Chern-Simones
coupling leads to the fact that baryon can be treated as a delta
function source of the gauge field $\mbox{${\cal A}$}_0 $ of brane [36].\\
The phenomenological quantities of baryon at finite temperature are
interesting topics, but the calculations are so complicated even in
lattice QCD. One can use the dual string theory to analyze most of
these concepts. The baryon binding energy, baryon melting
temperature and screening length are some of these examples which
are investigated using the $AdS_5 \times S^5$ dual string [27,37].
Furthermore some baryon thermodynamical quantities such as the
energy density and pressure have been studied in $SS$ model. But the
baryon binding energy in $SS$ model have not considered
yet and in this paper we plan to analyze it.\\
In section 2 we review the $SS$ model briefly in which the baryon is
considered as a $D_4$-brane wrapped on $S^4$ . In section  3 we use
the $U(1)_v$ field introduced by Ref. [32-35] to calculate the
energy of baryon configuration in $SS$ model. In section 4 we
subtract the dissociated baryon energy and obtain an equation for
the baryon binding energy in $SS$ model. Next we plot the energy in
terms of radial coordinate and then determine the stability range
for baryon configuration in $U_{KK}=1$ scale. We also obtained the
position and energy of the stable equilibrium point by using this
diagram. Then we depict the binding energy of baryon versus
temperature and estimate the critical $T$ for baryon melting.

\section{\textbf{$SS$} model}

In this section we review of Sakai-Sugimoto model with
$D_4/D_8/\overline{D_8}$ configuration [32]. The $D_4$-brane metric
is given by the following equation,
\begin{eqnarray}
&&ds^2=\left(\frac{U}{R}\right)^{3/2} \left(\eta_{\mu\nu}dx^\mu
dx^\nu+f(U)d\tau^2\right) +\left(\frac{R}{U}\right)^{3/2}
\left(\frac{dU^2}{f(U)}+U^2 d\Omega_4^2\right),
\nonumber\\
&&~~e^\phi= g_s \left(\frac{U}{R}\right)^{3/4},
~~~~F_4=dC_3=\frac{2\pi N_c}{V_4}\epsilon_4 \  , ~~~~~f(U)=1-\frac{\
U_{\rm KK}^3}{U^3}, \ \label{Eq 1}
\end{eqnarray}
where $D_4$-brane is extended along the $x^\mu (\mu=0,1,2,3)$ and
$\tau$ directions. $U (U \geq U_{KK})$ is the radial coordinate and
$d\Omega_4^2 $, $\epsilon_4$ and $V_4=8\pi^2/3$ are line element,
volume form and the volume of $S^4$. $R$ and $U_{KK}$ are constant
parameters.\\
Avoiding to have a singularity in $U=U_{KK}$, the $\tau$ coordinate
should be considered periodic,
\begin{eqnarray}
\tau\sim\tau+\delta\tau \ ,~~~~\delta\tau\equiv\frac{4\pi}{3}
\frac{R^{3/2}}{\ U_{\rm KK}^{1/2}} \ . \label{Eq 3 }
\end{eqnarray}
The $R$ and $U_{\rm KK}$ parameters are defined in terms of $l_s$
and $\lambda(=g_{Y\!M}^2 N_c)$ as follows,
\begin{eqnarray}
R^3=\frac{1}{2}\frac{ \lambda l_s^2}{\ M_{\rm KK}} \ ,~~~ \ U_{\rm
KK}=\frac{2}{9}\lambda M_{\rm KK} l_s^2 \ ,~~~
g_s=\frac{1}{2\pi}\frac{\lambda}{M_{\rm KK} l_s N_c} \ ,\label{Eq 4
}
\end{eqnarray}
and also the pion decay constant has the following expression,
\begin{eqnarray}
f_\pi^2=\frac{1}{54\pi^4}(g_{Y\!M}^2 N_c) M_{\rm KK}^2 N_c .
\label{Eq 5 }
\end{eqnarray}
according to equation (1) for $N_f$  $D_8$-brane placed in $D_4$
background, equation (1), the action can be written by the follows,
\begin{eqnarray}
S_{D8}&=& S^{DBI}_{D8} + S^{CS}_{D8}\ ,
\nonumber \\
S^{DBI}_{D8}&=& -T_8 \int d^9 x \ e^{-\phi}\ {\rm tr}\,
\sqrt{-\det(g_{MN}+2\pi\alpha' F_{MN})} \ , \nonumber \\
S^{CS}_{D8}&=&\frac{1}{48\pi^3} \int_{D8} C_3 {\rm tr}\, F^3  \
,\label{Eq 6 }
\end{eqnarray}
where $T_8=\frac{1}{2\pi^8\,l_s^9}$ is the $D_8$-brane tension,
$F_{\rm MN}= \partial_M A_N - \partial_N A_M -i[A_M,A_N],
(M,N=0,1,..8)$ is the field strength tensor and $A_M$ is the
$U(N_f)$ gauge field on $D_8$-brane. The induced metric on
$D_8$-brane is given,
\begin{eqnarray}
ds^2=\left(\frac{U}{R}\right)^{3/2}\!\!\!\!  \eta_{\mu\nu}dx^\mu
dx^\nu + \left[ \left(\frac{U}{R}\right)^{3/2}\!\!\!\! f(U)
(\tau'(U))^2 +\left(\frac{R}{U}\right)^{3/2}\!\!\!\! \frac{1}{f(U)}
\right]dU^2+ \left(\frac{R}{U}\right)^{3/2}\!\!\!\! U^2 d\Omega_4^2
\ ,\label{Eq 7}
\end{eqnarray}
where $U=U(\tau)$.\\
Baryon is a $D_4$-brane wrapped on $S^4$ in $SS$ model. In the other
hand, it is a soliton in Skyrme model and these two descriptions are
related to each other. The topological charge carried by the gauge
configuration on $D_8$-brane is related to the baryon number charge
and the skyrmion constructed on $D_8$-brane relates to the wrapping
$D_4$-brane. The relation between instanton number and $D_4$-brane
charge is [32],
\begin{eqnarray}
\frac{1}{8\pi^2} \int_B {\rm tr}\, F^2 = n ,\label{Eq 8}
\end{eqnarray}
where $B \simeq R^4$ is the four dimensional space parameterized by
$(x^1,x^2,x^3,z)$.\\
Inserting the appropriate gauge field in Chern-Simones action, one
can write,
\begin{eqnarray}
S_{CS}^{D8}\simeq n N_c \int_{R}  a , \label{Eq 9}
\end{eqnarray}
where $a$ is the $U(1)_\texttt{v}$ gauge field fluctuations on
$D_8$-brane.

\section{The energy of baryon in \textbf{$SS$} model}

In this section we calculate the energy of baryon configuration in
$SS$ model. The $CS$ coupling in $SS$ model leads to the fact that
instanton configuration displays a particle with $U(1)_v$ charge
$nN_c$ which is equivalent to a particle with baryon number $n_B$.
So we consider the baryon as a delta function source of the gauge
field [36]. The baryon action can be regarded as sum of the $DBI$
action and the source action. The $DBI$ action for $D_8$-brane in
the absence of the source term is written by the following equation,
\begin{equation}
S_{\textrm{D8}} =
 -\frac{N_f T_8 V_4}{g_s} \int d^4x
\, dU U^4 \left[ f \, (\tau')^2 + \left( \frac{R}{U} \right)^3
\left( f^{-1} - \left( 2\pi\alpha' \mbox{${\cal A}$}_{0}' \right)^2
\right) \right]^{\frac{1}{2}}, \label{Eq 10}
\end{equation}
where ${\cal A}_0'=\frac{d {\cal A}_0}{dU}$ and the $CS$ term
vanishes. From the equation of motion for $\tau(U)$ only the
$\tau'=0$ case is considered [32]. It means that the existence of
${\cal A}_0$ does not change the $D_8$-brane configurations in
$\tau$ coordinate. And it corresponds to $\tau = \frac{\delta
\tau}{4}$ which is the maximum of asymptotic separation between
$D_8$ and $\overline{D_8}$.\\
In order to express our results according to the $SS$ model we
choose  $z$ instead of $U$ as,
\begin{eqnarray}
U \equiv (U_{\rm KK}^3 + U_{\rm KK} z^2)^{1/3}, \label{Eq 11}
\end{eqnarray}
and then use the following dimensionless parameters,
\begin{eqnarray}
Z \equiv \frac{z}{U_{\rm KK}}, \quad K(U) \equiv 1+ Z^2 =
\left(\frac{U}{U_{\rm KK}}\right)^3 \ .\label{Eq 12}
\end{eqnarray}
Then, one can rewrite the action (9) in the following form,
\begin{eqnarray}
S_{\textrm{D8}} =  -a \int d^4x \int dZ\, K^{2/3}  \, \sqrt {\, 1 -
b K^{1/3} (\partial_Z \mbox{${\cal A}$}_{0})^2 } , \label{Eq 13}
\end{eqnarray}
where
\begin{eqnarray}
a \equiv \frac{N_c N_f \l_s^3 M_{\rm KK}^4}{3^9 \pi^5} \ , \quad
\quad b \equiv \frac{3^6 \pi^2}{4 \l_s^2 M_{\rm KK}^2} \ .\label{Eq
14}
\end{eqnarray}
In the other hand, the source action has the following form [36],
\begin{eqnarray}
S_{\mathrm{source}}= N_c n_B \int d^4x \int dZ\, \delta
(Z)\mbox{${\cal A}$}_0(Z) .\label{Eq 15}
\end{eqnarray}
where $N_c n_b=n_q$ is the quark density.\\
Now by using the equations (12) and (14), we write the baryon action
as follows,
\begin{eqnarray}
S_{\textrm{Baryon}} =  &-&a \int d^4x \int dZ\, K^{2/3}  \, \sqrt
{\,
1 - b K^{1/3} (\partial_Z \mbox{${\cal A}$}_{0})^2 }\nonumber \\
&+&N_c n_B \int d^4x \int dZ\, \delta (Z)\mbox{${\cal
A}$}_0(Z).\label{Eq 16}
\end{eqnarray}
At first, we should solve the equation of motion for the gauge field
to reach the baryon hamiltonian,
\begin{eqnarray}
\frac{d }{dZ} \frac{\partial {\cal L}}{\partial (\partial_Z
\mbox{${\cal A}$}_0)}
 = n_q \delta(Z)\ , \label{Eq 17}
\end{eqnarray}
By the definition $D =\frac{\partial {\cal L}}{\partial (\partial_Z
\mbox{${\cal A}$}_0)} $ and integrating over $z$, the equation of
motion takes the following form,
\begin{eqnarray}
D = \frac{1}{2} n_q \ \mathrm{sgn}(Z)\ . \label{Eq 18}
\end{eqnarray}
In that case the equation (18) helps us to obtain the corresponding
baryon action and energy. We utilize the definition of $D$ to
eliminate $\partial_Z \mbox{${\cal A}$}_0$ as following,
\begin{eqnarray}
(\partial_Z \mbox{${\cal A}$}_0)^2=\frac{D^2}{a^2b^2K^2+bD^2K^{1/3}}
. \label{Eq 19}
\end{eqnarray}
Inserting this equation into the baryon action (16) yields to,
\begin{eqnarray}
S_{\textrm{Baryon}} =  &-& 2a^2 b \int d^4x \int dZ\, K^{5/3}  \,
(\, 4
a^2 b^2 K^{2}+ n_q^2 b  K^{1/3} )^{-1/2}\\
\nonumber &+& n_q \int d^4x \int dZ\, \delta (Z)\mbox{${\cal
A}$}_0(Z),\label{Eq 20}
\end{eqnarray}
which is independent of $\partial_Z \mbox{${\cal A}$}_0$.\\
Then to obtain the baryon energy we should transform the original
lagrangian to eliminate the gauge field in favor of $D$ as follows,
\begin{eqnarray}
\mbox{${\cal L}$}_{\textrm{Baryon}} \rightarrow  n_q \delta
(Z)\mbox{${\cal A}$}_0(Z) - \mbox{${\cal L}$}_{\textrm{Baryon}}
.\label{Eq 21}
\end{eqnarray}
After this transformation, the baryon lagrangian will be following,
\begin{eqnarray}
\mbox{${\cal L}$}_{\textrm{Baryon}} = 2a^2 b K^{5/3}  \, (\, 4 a^2
b^2 K^{2}+ n_q^2 b  K^{1/3} )^{-1/2}.\label{Eq 22}
\end{eqnarray}
Substituting this equation into the baryon action (19) we find the
energy of baryon configuration as following,
\begin{eqnarray}
\mbox{${E}$}_{\textrm{Baryon}} = 2a^2 b V_3 \int dZ K^{5/3} \, (\, 4
a^2 b^2 K^{2}+ n_q^2 b  K^{1/3} )^{-1/2}, \label{Eq 23}
\end{eqnarray}
where $V_3$ is the spatial integral. Here we attain an equation for
the baryon energy assuming the baryon as a delta function in terms
of the $SS$ model parameters.

\section{Baryon binding energy}

In this section we obtain the baryon binding energy with a good
approximation. For this purpose we should subtract the energy of dissociated
baryon from the total energy of baryon configuration (equation (21)).\\
By using the following relation the energy of dissociated baryon is
considered as the mass of $N_c$ deconfined quarks at the black hole
horizon,
\begin{eqnarray}
E_{diss} = \frac{N_c }{2 \pi \alpha'} \int \, dU  . \label{Eq 24}
\end{eqnarray}
where we assumed $\phi=0$ for the dilaton. \\
Then we have applied the equations (10) and (11) and rewrite the
equation (24) in terms of new dimensionless variable $Z$ as follows,
\begin{eqnarray}
E_{diss} = c \int \, dZ \, Z \, K^{-2/3}  , \label{Eq 25}
\end{eqnarray}
where
\begin{eqnarray}
c = \frac{N_c \, U_{KK}}{3 \pi \alpha'}  . \label{Eq 26}
\end{eqnarray}
Subtracting this equation from the baryon energy (equation (22)),
one can obtain the following equation for the baryon binding energy,
\begin{eqnarray}
E_{I} = 2a^2 b V_3 \int \, dZ \, K^{5/3} \, (\, 4 a^2 b^2 K^{2}+
n_q^2 b K^{1/3} )^{-1/2} - c \int \, dZ \, Z \, K^{-2/3}. \label{Eq
27}
\end{eqnarray}
The first integrate can not be solved analytically but we have to
consider the following condition which is obtained by numerical
values for the parameters,
\begin{eqnarray}
n_B\,<\, \frac{N_f \, \lambda^2 \, M_{KK}^3}{3^6 \, \pi^4}.
\label{Eq 28}
\end{eqnarray}
which implies that $n_q<\,2\,a\,b^{1/2}$ and simplifies the
integrand. Thus the baryon binding energy is approximated up to the
second order of the power expansion,
\begin{eqnarray}
E_{I} = a V_3 \int dZ K^{2/3} \,
\{\,1-\frac{1}{2}\frac{n_q^2}{2a^2bK^5/3} +... \} - c \int \, dZ \,
Z \, K^{-2/3}, \label{Eq 29}
\end{eqnarray}
which has analytical solution in terms of the hypergeometric
functions as follows,
\begin{eqnarray}
E_{I} &=&  \, a \, V_3 \,  Z \,
F([\frac{-2}{3},\frac{1}{2}],[\frac{3}{2}],-Z^2) \, \\
\nonumber &-& \,\frac{1}{2} c\, Z^2 \,F([\frac{2}{3},1],[2],-Z^2)
-\frac{1}{8} \,\frac{V_3\,n_q^2 }{a\,b} \, tan^{-1}(Z) . \label{Eq
30}
\end{eqnarray}
This is the baryon binding energy in terms of the dimensionless
parameter $Z$ introduced in $SS$ model.\\
Finally we change $Z$ into $U$, the radial coordinate in $SS$ model,
and plot the baryon binding energy in terms of this coordinate.
Then, we use this plot to determine the range of $U$ in which the
baryon is stable in $SS$ model. We also appraise the position and
energy of the stable equilibrium point. The equation (27) is applied
for baryon density and the numerical value of $\lambda$ is obtained
from equation (4) with $f_\pi = 0.093\,GeV$ for the experimental
value of the pion decay constant. Also we choose $N_f=2$ and $N_c=3$
values and insert these values with $M_{kk}=0.950 \,GeV$ in equation
(29). Then we plot the baryon binding energy with respect to the $U$
coordinate in $U_{kk}= 1$ scale (figure 1-a). As $U$ increases, the
binding energy of baryon configuration gets smaller and at $U=U_b$
we have an stable equilibrium point with the minimum energy of $E_I
= -2.36\, GeV$. At $U = U_m$ the binding energy is zero and for $U\,>\,U_m$ the baryon would be dissociated.
So we obtain an stable range for the baryon configuration.\\
Furthermore, in equation (29) we use the relation between the
horizon coordinate $U_{KK}$ and the temperature to depict the baryon
binding energy in terms of $T$ at $U\,=\,U_m$ (figure 1-b). As $T$
increases, the baryon binding energy becomes larger and for $T>T_c$
we have no stable configuration. Note that we obtained a similar
behavior for the binding energy versus $T$ compared to Ref.[37] with
the $AdS_5\times S^5$ configuration.
\begin{figure}[bth]
\centerline{\includegraphics[width=14cm]{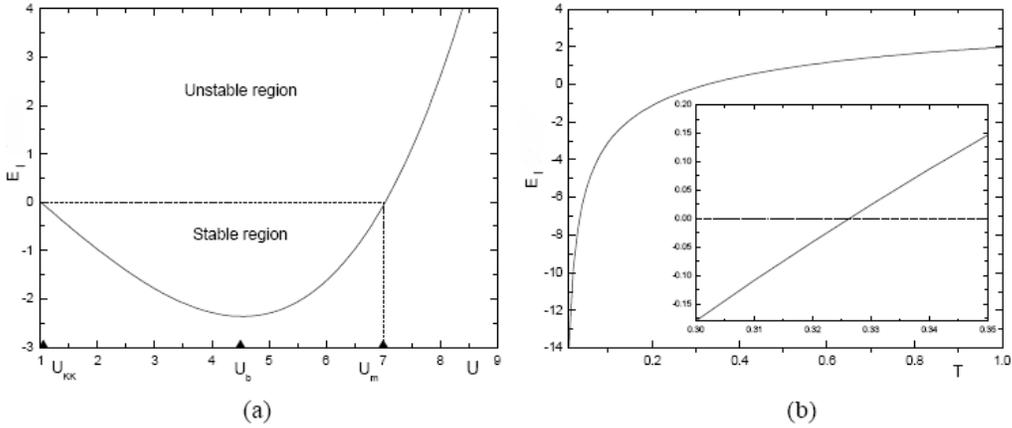}} \caption{a)
The baryon binding energy vs. $U$ coordinate in $SS$ model. The
baryon is stable in $ U_{kk} < U < U_m $ range and $ U = U_b $ is
the stable equilibrium point. b) The baryon binding energy vs. $T$.
$T_c$ is the temperature in which the binding energy becomes zero
and for $T>T_c$ no stable configuration exists. \label{fig1}}
\end{figure}
\\
\section{Conclusion}
We calculated the binding energy of baryon in the gauge/gravity dual
description with a $D_4/D_8/\overline{D_8}$ introduced in $SS$
model. Here we considered the baryon action as the sum of the $DBI$
action and a delta function source of the gauge field as in [36]. We
obtained the total energy of baryon and the energy of $N_c$
fundamental quarks in section 3. Then in order to obtain the baryon
binding energy, we subtracted the energy of dissociated baryon from
the total energy of baryon. Finally we depicted the baryon binding
energy versus the radial coordinate and determined the stability
range for baryon configuration by using this diagram. Also we
plotted the baryon binding energy versus the temperature.
Then the critical temperature $T_c$ can be realized clearly.\\
According to the binding energy graph we can easily find that the
baryon binding energy is zero at $U=U_{kk}$ which is the lower bound
for $U$ coordinate in $SS$ model where the radius of $S^{1}$
diminishes to zero and no stable baryon configuration exists. As $U$
increases, the binding energy of baryon configuration gets smaller
and at $U=U_b$ we have an stable equilibrium point. For $U > U_b$
the energy increases and at $U = U_m$ the binding energy vanishes
again. It reveals the fact that for $U > U_m$ there is no stable
baryon configuration and the baryon would be dissociated.
Furthermore we obtained a similar behavior for the binding energy
versus $T$ compared to Ref.[37] which proposes an $AdS_5\times S^5$
configuration.



\begin{thebibliography}{99}
\bibitem{1} Maldacena J M 1998 The large N limit of superconformal field theories and supergravity Adv. Theor. Math.
Phys. 2 231.
\bibitem{2}Witten E 1998 Anti-de Sitter space and holography Adv. Theor. Math. Phys. 2
253.
\bibitem{3}Schwart J H 1999 Introduction to M Theory and AdS/CFT Duality (Lecture Notes in Physics vol 525) (Berlin:
Springer) pp 1–21 (arXiv:hep-th/9812037).
\bibitem{4}Douglas M R and Randjbar-Daemi S 1999 Two lectures on AdS/CFT correspondence
arXiv:hep-th/9902022.
\bibitem{5}Petersen J L 1999 Introduction to the Maldacena conjecture on AdS/CFT Int. J. Mod. Phys. A 14
3597.
\bibitem{6}Nastase Horatiu 2007 Introduction to AdS-CFT arXiv:0712.0689v2
[hep-th].
\bibitem{7}Klebanov Igor R 2000 TASI lectures: introduction to the AdS/CFT correspondence
arXiv:hep-th/0009139.
\bibitem{8}Sadeghi J, Setare M R, Pourhassan B 2009 Drag force with different charges in STU background
and AdS/CFT J. Phys. G: Nucl. Part. Phys. 36 (2009) 115005 (19pp).
\bibitem{9} Sadeghi J, Setare M R, Pourhassan B and Hashmatian S 2009 Drag force of moving quark
in STU background Eur. Phys. J. C 61 527 (arXiv:0901.0217 [hep-th]).
\bibitem{10}Sadeghi J and Pourhassan B 2008 Drag force of moving quark at the N = 2 supergravity J. High Energy Phys. JHEP12(2008)026
(arXiv:0809.2668 [hep-th]).
\bibitem{11}Herzog CP, Karch A, Kovtun P, Kozcaz C and Yaffe L G 2006 Energy loss of a heavy quark moving through
N = 4 supersymmetric Yang–Mills plasma J. High Energy Phys.
JHEP07(2006)013 (arXiv:hep-th/0605158).
\bibitem{12}Herzog C P 2006 Energy loss of heavy quarks from asymptotically AdS geometries J. High Energy Phys.
JHEP09(2006)032 (arXiv:hep-th/0605191).
\bibitem{13}Gubser S S 2006 Drag force in AdS/CFT Phys. Rev. D 74 126005.
\bibitem{14}Vazquez-Poritz J F 2008 Drag force at finite 't Hooft coupling from AdS/CFT
arXiv:hep-th/0803.2890.
\bibitem{15}Caceres E and Guijosa A Drag force in charged N = 4 SYM plasma J. High Energy Phys.
JHEP11(2006)077.
\bibitem{16}Matsuo T, Tomino D andWenWY 2006 Drag force in SYM plasma with B field from AdS/CFT J. High Energy
Phys. JHEP10(2006)055.
\bibitem{17}Nakano E, Teraguchi S and Wen W Y 2007 Drag force, jet quenching and
AdS/QCD Phys. Rev. D 75 085016.
\bibitem{18}Liu H, Rajagopal K and Wiedemann U A 2006 Calculating the jet quenching parameter from AdS/CFT Phys.
Rev. Lett. 97 182301.
\bibitem{19}Vazquez-Poritz J F 2006 Enhancing the jet quenching parameter from marginal deformations arXiv:
hep-th/0605296.
\bibitem{20}Caceres E and GuijosaA2006 On drag forces and jet quenching in strongly coupled plasmas J. High Energy Phys.
JHEP12(2006)068.
\bibitem{21}Lin F L and Matsuo T 2006 Jet quenching parameter in medium with chemical potential from AdS/CFT Phys.
Lett. B 641 45.
\bibitem{22}Avramis S D and Sfetsos K 2007 Supergravity and the jet quenching parameter in the presence of R-charge
densities J. High Energy Phys. JHEP01(2007)065.
\bibitem{23}Armesto N, Edelstein J D and Mas J 2006 Jet quenching at finite 't Hooft coupling and chemical potential from
AdS/CFT J. High Energy Phys. JHEP09(2006)039.
\bibitem{24}Edelstein J D and Salgado C A 2008 Jet quenching in heavy Ion collisions from AdS/CFT AIP Conf. Proc. 1031
207 (arXiv:0805.4515).
\bibitem{25}Fadafan K B 2008 Medium effect and finite t'Hooft coupling correction on drag force and jet quenching
parameter arXiv:0809.1336.
\bibitem{26}Peeters K, Sonnenschein J and Zamaklar M 2006 Holographic melting and related properties of mesons in a
quark gluon plasma Phys. Rev. D 74 106008.
\bibitem{27}Liu H, Rajagopal K and Wiedemann U A 2007 An AdS/CFT calculation of screening in a hot wind Phys. Rev.
Lett. 98 182301 (arXiv:hep-ph/0607062).
\bibitem{28}Chernicoff M, Garcia J A and Guijosa A 2006 The energy of a moving quark-antiquark pair in an N = 4 SYM
plasma J. High Energy Phys. JHEP09(2006)068.
\bibitem{29}Erdmenger J, Evans N, Kirsch I and Threlfall E J 2008 Mesons in gauge/gravity duals Eur. Phys. J. A 35
81.
\bibitem{30}Sadeghi J, Pourhassan B and Heshmatian S 2008 Rotating heavy meson in an N = 4 SYM plasma and AdS/CFT
arXiv:0812.4816 [hep-th]; Ali-Akbari M and B. Fadafan K 2009 Rotating mesons in the presence
of higher derivative corrections from gauge-string duality arXiv:0908.3921v1 [hep-th].
\bibitem{31}Chernicoff M, Garcia J A and Guijosa A 2006 The energy of a moving quark-antiquark pair in an N = 4 SYM
plasma J. High Energy Phys. JHEP09(2006)068.
\bibitem{32}Sakai T and Sugimoto S 2004 Low Energy Hadron Physics in Holographic QCD arXiv:0412141v5 [hep-th].
\bibitem{33}Sakai T and Sugimoto S 2005 More on a Holographic Dual of QCD arXiv:0507073v4 [hep-th].
\bibitem{34}Hashimoto K Sakai T and Sugimoto S 2008 Holographic Baryons arXiv:0806.3122v3 [hep-th].
\bibitem{35}Hata H Sakai T Sugimoto S and Yamato S 2007 Baryons from instantons in holographic
QCD arXiv:0701280 [hep-th].
\bibitem{36} Kim K Y, Sin S J and Zahed I 2007 The Chiral Model of Sakai-Sugimoto at Finite Baryon
Density arXiv:0708.1469v3 [hep-th].
\bibitem{37}Sin S J, Yang S and Zhou Y 2009 Comments on Baryon Melting in Quark Gluon Plasma
with gluon condensation arXiv:0907.1732v1 [hep-th].



\end{thebibliography}
\end{document}